\documentclass[12pt]{article}
\usepackage{graphicx}
\usepackage{amsmath}
\usepackage{tabularx}
\usepackage{float}
\usepackage{a4wide}
\usepackage{mathpazo}
\usepackage{authblk}

\usepackage[
    font=scriptsize,
    labelfont=bf,
    ]{caption}

\usepackage[
        backend=biber,
        sorting=none,
        style=numeric
    ]{biblatex}

\usepackage{xcolor}
\usepackage{amsfonts}
 
\usepackage{enumitem}

\usepackage{etoolbox}
\def\enum{\begin{newEnumerate}}
\def\endenum{\end{newEnumerate}}

\title{The origins of large-scale structure in family networks}

\author[1]{Lasse Mohr}
\author[2,3]{Andreas Bjerre-Nielsen}
\author[1,3]{Sune Lehmann}
\affil[1]{Department of Applied Mathematics and Computer Science, Technical University of Denmark}
\affil[2]{Department of Economics, University of Copenhagen}
\affil[3]{Center for Social Data Science (SODAS), University of Copenhagen}

\begin{document}
\maketitle

\begin{abstract}
Family relations are the most fundamental of all social networks and encompass everyone.
Family networks grow as individuals have children, creating connections between families, which over time create large and complex structures.
While partner-choice homophily has been proposed as a key driver in this growth process, little is known about the connection between individual behavior and the emergent large-scale structure of family networks.
Here, we analyze a unique population-complete family network, covering millions of individuals across several decades, enriched with demographic, educational, and geographic data from high-quality national registries.
Drawing on the longitudinal coverage of our observations and using a series of growing-network models, we unravel how individual-level behavior shapes the large-scale network structure.
Contrary to prevailing theories, we find that partner-choice homophily has little effect on the emergent large-scale structure.
Instead, we identify two key drivers: 
First, partner-change behavior, where individuals leave one partner for another, creates `shortcuts' in the network akin to rewirings in the Watts-Strogatz model.
These shortcuts decrease pathlengths and accelerate the emergence of meso-scale connected components. 
Second, we find that partner change is a self-exciting behavior, such that the probability of changing partner increases with an individual's prior number of partners.
The self-exciting behavior accelerates the generation of large network components, with highly connected individuals functioning as network hubs.
Accounting for this partner-change behavior, we are able to accurately capture multiple large-scale network properties of the empirical family network.
Finally we show that homophily-driven behavior is not able to generate the observed network structure.
\end{abstract}


%%%%%%%%%%%%%%%%%%%%%%%%%%%%%%%%%%%%%%%%%%%%%%%%%%%%%%%%%%%%%%%%%%%%%%%%%%%%%%%%%%%%%%%%%%%%%%%%%%%%%
\section*{Introduction}
%%%%%%%%%%%%%%%%%%%%%%%%%%%%%%%%%%%%%%%%%%%%%%%%%%%%%%%%%%%%%%%%%%%%%%%%%%%%%%%%%%%%%%%%%%%%%%%%%%%%%

Networks of family relations are central in the social fabric of societies and emerge from individual fertility decisions, such as whether or not to have children, how many children to have, and with whom.
Fertility decisions shape individual families and govern how families connect to create larger structures.
However, this connection between individual behavior and large-scale network structure has never been directly studied, leaving the structural consequences of different behaviors unclear.

The social factors driving the choice of romantic partners, also known as `assortative mating', has been studied in economics and sociology~\cite{LuoAssortative2017, schwartzAssortative2009, SchwartzEducation2010, GreenwoodMaryYourLike2014, rauscherInfantHealth2020, almarMarital2023, EikaEducation2019, Speakman2007, Alford2011, Horwitz2023, Yamamoto2023, PeyrotGenetic2016}, and has been assumed to be a fundamental mechanism governing structure when modeling processes on family networks~\cite{Collado2023, Yengo2018, Fernandez2001}.

Beyond how individuals with similar individual-level qualities tend to form couples, other aspects of family networks have been studied extensively across the sciences and across over 150 years~\cite{Galton1869}.
Researchers have considered biases and society-level consequences of partner choices with respect to income~\cite{EikaEducation2019, GreenwoodMaryYourLike2014, SchwartzEarning2010, schwartzTrends2013, ErmischSorting2006}, educational attainment~\cite{Collado2023, Fernandez2001}, and health~\cite{Derraik2015, PeyrotGenetic2016, rauscherInfantHealth2020};
trends and confounders of individual-level fertility~\cite{Hoem2006Childless, Lesthaeghe2014Transition, AhmadzadehTori2023, Diaz2011, Balbo2013Fertility, OECD2023AgeAtChildbirth, Beaujouan2019},
partner-change behavior~\cite{Thomson2020Multi-Partner, SchachtKramer2019, Kuang2025, ManloveEtAl2008};
the branching of ancestral lines~\cite{Chang1999CommonAncestor, rohde2004modelling};
local family network structure~\cite{AlburezGutierrez2023, Daw2016, Kolk2023, Salvatore2018Divorce, CaswellTwoSex2022, CaswellStochastic2024};
genetic heritability~\cite{Kaplanis2018, Stefansdottir2013,Morrison2016, Malmi2018, varga2023Genealogical, Colasurdo2024FamiLinx};
marriage networks~\cite{levi-strauss1969elementary, hamberger2011KinshipNetwork};
and the effect of family networks on long-term social mobility \cite{clark2014surnames, clark2023Inheritance, hallsten2023Shadow, collado2023extended}.
In spite of this deep history, the drivers of the emergent large-scale structure of family networks have never been quantified.

Drawing on a completely unique dataset~\cite{Nielsen2024NationScale, Cremers2025Temporal} covering the entire population of Denmark $\sim\,$6 million individuals and $\sim\,$8 million parent-child relations (all such family relations from 1953 to 2018), our study empirically investigates the link between individual-level behavior and large-scale structure with the aim of quantifying the relative importance of partner-choice homophily as compared to other behavioral traits.
The number of relations in our dataset is three orders of magnitudes larger than any other family network studied to date~\cite{Boyd2023, hamberger2011KinshipNetwork}.
The deep longitudinal span and complete population coverage of this network is what allow us to model how individual-level behavior aggregates to become large-scale structure (Appendix \ref{app:danish_family_network}).

Different criteria for when a pair of individuals constitute a couple have been used in the literature, including marriage~\cite{schwartzTrends2013}, dating~\cite{BruchDatingMarket2019, Abramova2016}, and cohabitation~\cite{sasslerCohabition2020}.
Here, because we seek to quantify the effects of partnering behavior on large-scale structure of family networks, we define couples as \textit{coparents}, pairs of individuals that have one or more children together. 
This explicitly links the behavior related to choice of partner to the family-network structure.

Our work reveals that while partner-choice homophily is important for describing the detailed matching patterns of couples across sociodemographic and geographic attributes, it only plays a minor role in shaping the large-scale network properties. 
Instead we find that large-scale structure is predominately determined by partner-change behavior, the patterns of how individuals in existing partnerships leave their partner for another (irrespective of individual-level qualities).
The intuitive reason that partner changes are crucial from a network perspective is that they create shortcuts in an otherwise very slowly growing and highly localized family network.
In this sense, partner changes function in a way very similar to that of shortcuts in the Watts-Strogatz small-world network model~\cite{Watts1998}.
We further find that partner changes tend to be self-exciting in the sense that an individual's probability of changing partner increases with previous-partner count (resembling preferential-attachment~\cite{Barabasi1999}).
This mechanism is novel and contributes to further accelerate the interconnectedness of the network.

In fact, partner changes and their self-exciting property, accurately generate many large-scale network traits observed in the empirical family network, such as the average shortest-path length, overall connectivity speed, percolation behavior, and the emergence of large connected components spanned by coparent relations.

Turning to the role of homophily in the network, we do find strong evidence of partner-choice biases with respect to age, length of education, geographic origin, and number of prior partners in terms of how pairs of individuals in the empirical network are connected.
However, including these partner-choice biases in our models reveals limited impact on the resulting overall network structure, only modestly improving our models' correspondence with the empirical network structure -- a negligible effect relative to what is achieved by including partner-change behavior.

Ours is the first study to measure and quantify the individual-level behavioral origins of the large-scale structure of family networks.

%%%%%%%%%%%%%%%%%%%%%%%%%%%%%%%%%%%%%%%%%%%%%%%%%%%%%%%%%%%%%%%%%%%%%%%%%%%%%%%%%%%%%%%%%%%%%%%%%%%%%
\section*{Results}
\subsection*{Partner change and family network structure}\label{sec:partner_change_counts}

Partner-change behavior plays a fundamental role in shaping the emergent structure of family networks.
A fundamental insights in the modern era of network science is that introducing shortcuts to an otherwise highly localized network radically changes its structural properties -- lowering the average shortest-path length~\cite{Watts1998} and change its percolation behavior~\cite{NewmanScaling1999}.
This phenomenon is commonly referred to as the `the small-world property'.

\begin{figure}[H]
    \centering
    \includegraphics[width=1.0\linewidth]{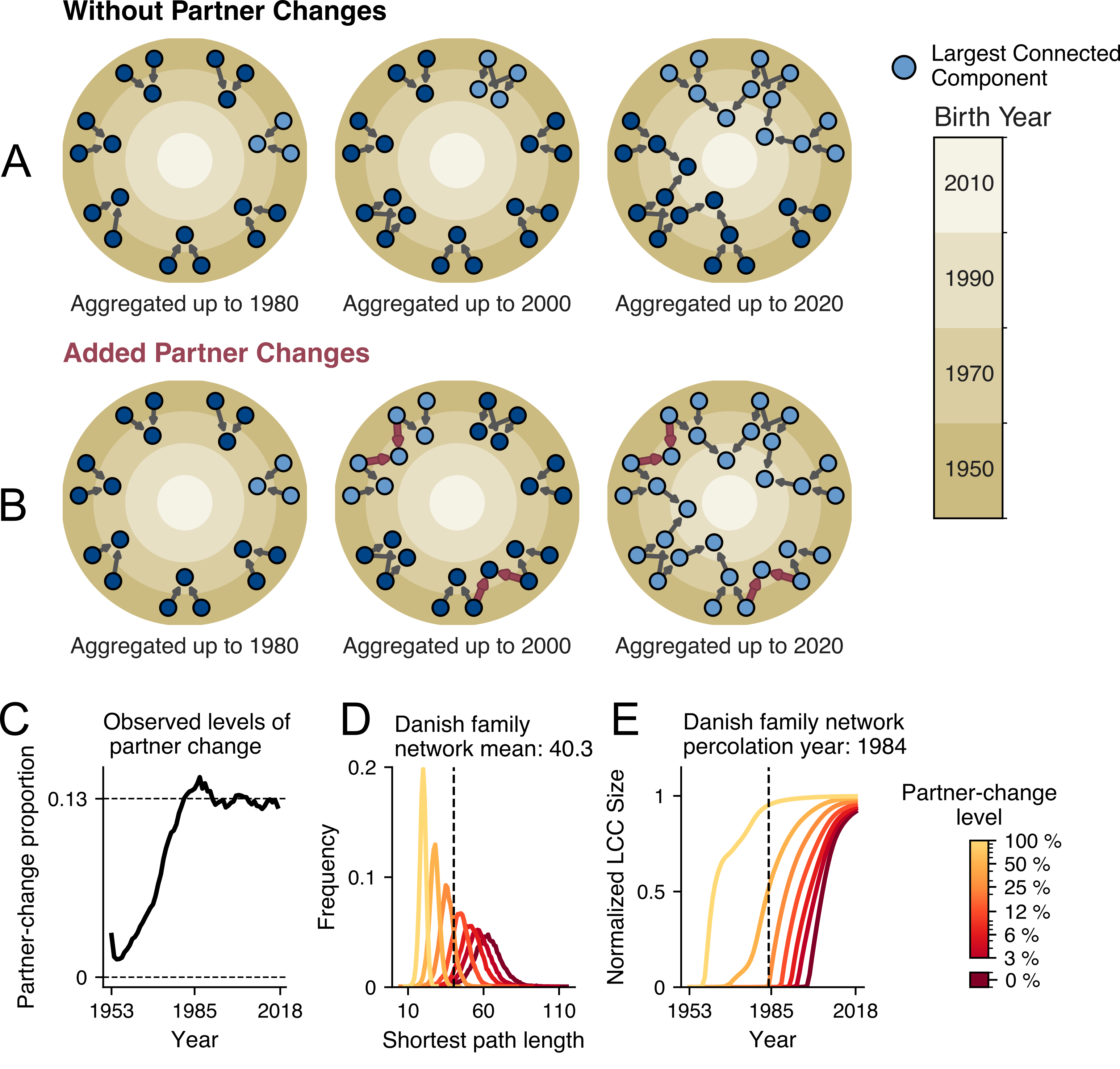}
    \caption{
        Family networks are highly localized with partner changes creating shortcuts across otherwise distant families. 
        (A) Growth of a toy family network in the absence of partner changes; families connect only once their children have children of their own. 
        As each family only has a small number of children, the resulting structure is localized and large components only emerge slowly.
        (B) Growth of a toy family network with partner changes.
        These connect otherwise distant or unconnected families, decreasing path-lengths and accelerating the emergence of large connected components.
        (C) Proportion of possible yearly partner-changes in the Danish family network.
        Partner-change stabilize around $13\%$.
        (D) Shortest undirected-path length distributions among uniform random node pairs of surrogate family networks generated for partner-change-levels, ranging from $0\%$ to $100\%$.
        (E) Growth the largest weakly-connected components of the same set of surrogate family networks.
        Networks with higher yearly partner-changes levels have lower average shortest-path length and earlier onset of percolation.
    }
    \label{fig:partner_change_counts_effect}
\end{figure}

\textbf{The role of shortcuts in the family network}.
A simple \textit{gedanken experiment} illustrates how this happens in the family network. 
Assuming observations are left-censored in 1950, a family network entirely without partner changes is initially made up of disconnected families each consisting of two parents and their child (Fig. \ref{fig:partner_change_counts_effect}\,A).
Aggregating over time, couples may have more children, but individual families remain disconnected for many years, simply because the children are not yet old enough to mature and reproduce.
Only when aggregating observations across several decades, do we observe connections between families.
Because each family can only connect to as many other families as they have children, these connections create a localized network with long path lengths.

In contrast, consider a family network where a set of individuals change partner.
This network initially appears identical to the one without partner changes (Fig. \ref{fig:partner_change_counts_effect}\,B).
However, individuals that change partners (highlighted in red) are able to create connections between families much sooner than any children reach adulthood.
Therefore, we only need to aggregate observations for short amount of time before connections form between families.
These early connections strongly impact the emerging structure as we continue to aggregate. 
This is exemplified by the size difference of the largest connected components of the final networks with and without partner changes for the toys networks in panel A and B of Figure \ref{fig:partner_change_counts_effect}.
From this analysis it is clear that when the majority of individuals only have a single partner, the resulting family network will be highly localized, and that individuals with multiple partners will form shortcuts, connecting their own and all their partners' families.

\textbf{A data-driven model for family networks}.
To empirically investigate the effects of partner-change behavior on the emerging network structure outlined above, and to later compare with the effects of partner-choice homophily, we formulate a generative random network model that controls for individual-level fertility and allows us to generate surrogate family networks with various degrees of partner change.
We outline the model below and provide full details in Appendix \ref{app:surrogate_family}.

To generate a network, we use the following procedure: for each person, we register which years they have one or more children, and denote the person an \textit{active parent} in those years.
Then, for each year of observations, we randomly match active parents into couples and assign each child born in the given year to a couple, controlling for sex and out-degree co-occurrences of couples. 
In terms of partner changes, we sample pre-specified numbers of active parents to change partner uniformly at random.
We then create directed edges from each child to both members of its assigned parent couple.
We refer to this procedure as the \textit{surrogate family-network model} and to the generated networks as \textit{surrogate family networks}.
By relying on reconfigurations of observed individuals instead of generating a hypothetical population through sampling, the surrogate family-network model avoids the known biases of other microsimulation models of family networks~\cite{RugglesConfessions1993}. 
Further, our models capture non-trivial traits of individual-level fertility such as dependency between individual age-at-first-child and out-degree, and correlation of out-degree between parents and children (see Appendix \ref{app:validate_model} for validation of the generated individual-level fertility traits).

\textbf{Shortest paths and percolation}.
In the Danish family network, we observe approximately $13\%$ of the total number of possible partner-changes (Fig.\,\ref{fig:partner_change_counts_effect}\,C), with partner-change levels being difficult to estimate until the 1980's due to left-censored observations (see Appendix \ref{app:measuring_partner_changes}).
This results in $10.3\%$ of parent-nodes in the Danish family network having children with multiple partners, which is comparable to reported levels of multi-partner fertility in other western countries~\cite{Thomson2020Multi-Partner}.

To elucidate the effect of yearly partner-change levels on the emergent structure of family networks, we generate surrogate family networks for a range of different yearly partner-change levels, from zero changes per year to the maximum possible number of changes.
As expected, the shortest undirected-path distance distributions of uniform random node pairs reveals a clear tendency where networks with a larger fraction of partner changes have smaller shortest-path lengths (Fig.\,\ref{fig:partner_change_counts_effect}\,D).
The Danish family network has a mean shortest-undirected-path length of $40.3$ (vertical dashed line in Fig.\,\ref{fig:partner_change_counts_effect}\,D), which is still large compared to traditional social networks~\cite{UganderAnatomy2011}.

To study the effect of partner-change levels on the speed with which the network grows connected, we measure the size of the largest weakly-connected component when aggregating observations over different lengths of time.
The networks initially start out as small disconnected components, with the largest connected component making up a negligible fraction of the network.
However, aggregating observations over time, a giant connected component emerges, which over time grows to encompass most of the network.
We observe a clear dependence between partner-change level and the time it takes until the giant connected component emerges in the resulting surrogate family networks, where the more partner changes occur, the faster the component emerges (Fig.\,\ref{fig:partner_change_counts_effect}\,E).
The empirical data undergoes a similar phase transition when aggregating observations for 31 years up to 1984.

\begin{figure}[H]
    \centering
    \includegraphics[width=1\linewidth]{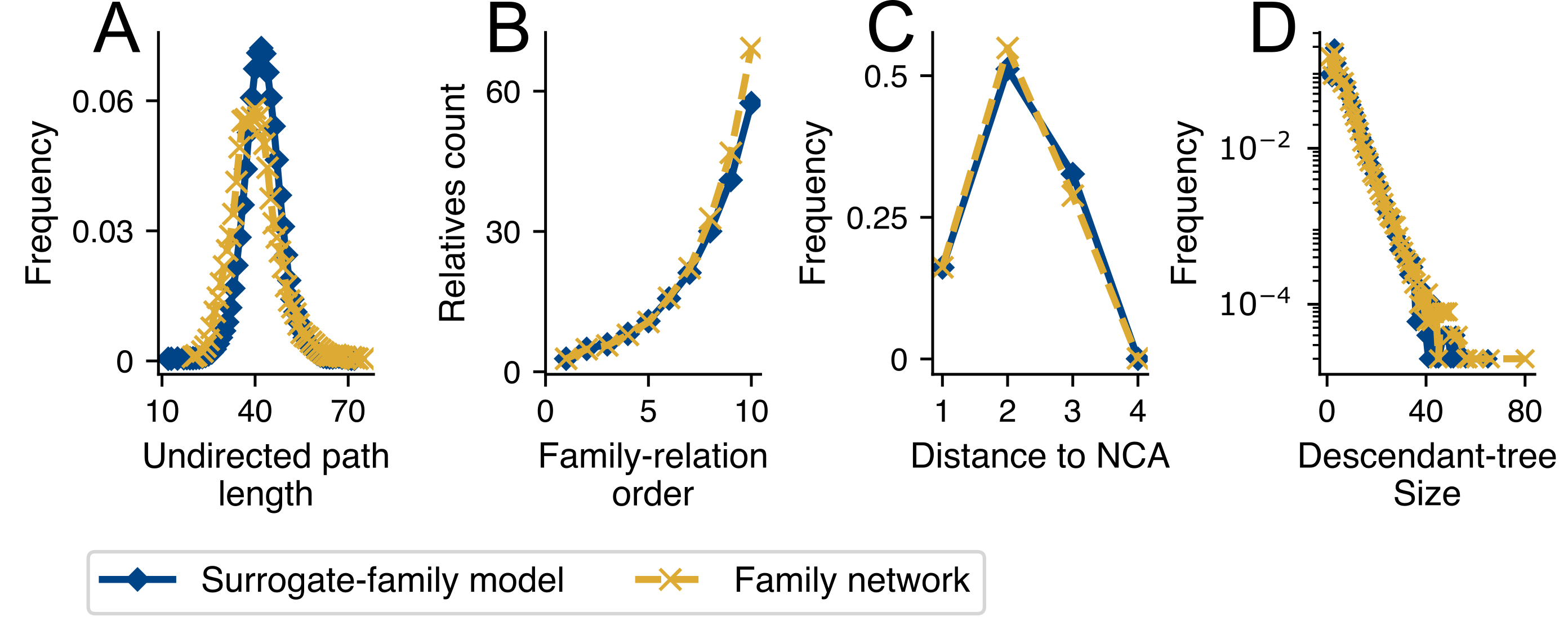}
    \caption{
    Accounting for yearly partner-change levels explains multiple emergent large-scale structural traits of family networks:
    (A) Shortest undirected-path length distribution; (B) average numbers of relatives across different orders (measured by path-length distance); (C) size distribution of descendant trees across source nodes; and (D) distance to nearest-common-ancestor distribution across leaf-pairs.
    All panels show the empirical statistics of the Danish family network and of the surrogate family-network model accounting for the empirical yearly partner-change level, averaged over 10 independent simulations.
    We observe that the these non-trivial large-scale traits of the emergent structure, the surrogate family networks mimic the Danish family network closely.
    }
    \label{fig:simple_model}
\end{figure}

\textbf{Our model accurately captures many aspects of network structure}.
Even though the surrogate family model is an extreme simplification of the underlying network growth process, simply by controlling for the observed yearly partner-change level, we find that the models accurately capture several properties of the emerging large-scale structure in the Danish family network.
For example, the shortest-undirected path length distributions closely match that of the Danish family network (Fig \ref{fig:simple_model} panel A), with a slight bias towards longer paths.

The kinship structure around individuals is often summarized as the average number of relatives per person~\cite{AlburezGutierrez2023, Daw2016, Kolk2023}.
Relatives can be stratified by their distance from the center node along a shortest path in the family network, which we denote by their relation order (with parents and children having order 1; grandparents, siblings, and grandchildren order 2, and so forth).
We find that the average numbers of relatives per person for the surrogate family networks matches those for the Danish family network for all orders up to $10$, at which point we observe a small discrepancy, (Fig.\,\ref{fig:simple_model}\,B).

Because family networks are directed acyclic graphs, each node is the source of a tree of descendants.
The descendants-count distribution for source nodes of the surrogate family networks accurately resemble that of the Danish family network (Fig.\,\ref{fig:simple_model}\,C).

Another important structure of family networks is the level of relatedness among pairs of nodes, which we measure as the minimum distance to a nearest-common-ancestor (NCA).
The NCA distance distribution amongst pairs of leaf nodes for the surrogate family networks also resemble that of the Danish family network (Fig.\,\ref{fig:simple_model}\,C).

%%%%%%%%%%%%%%%%%%%%%%%%%%%%%%%%%%%%%%%%%%%%%%%%%%%%%%%%%%%%%%%%%%%%%%%%%%%%%%%%%%%%%%%%%%%%%%%%%%%%%
\subsection*{Partner change behavior is heterogeneous}\label{sec:self-exciting}
%%%%%%%%%%%%%%%%%%%%%%%%%%%%%%%%%%%%%%%%%%%%%%%%%%%%%%%%%%%%%%%%%%%%%%%%%%%%%%%%%%%%%%%%%%%%%%%%%%%%%

By controlling for the observed partner-change level, the surrogate family-network model generate networks that share many traits of the empirical large-scale structure.
However, the simplest model fails to account for important micro-structures of partner-change behavior, which compound to affect the emergent \textit{coparent network}; the network of parent nodes with edges between nodes that have one or more children together.

\begin{figure}[H]
    \centering
    \includegraphics[width=0.8\linewidth]{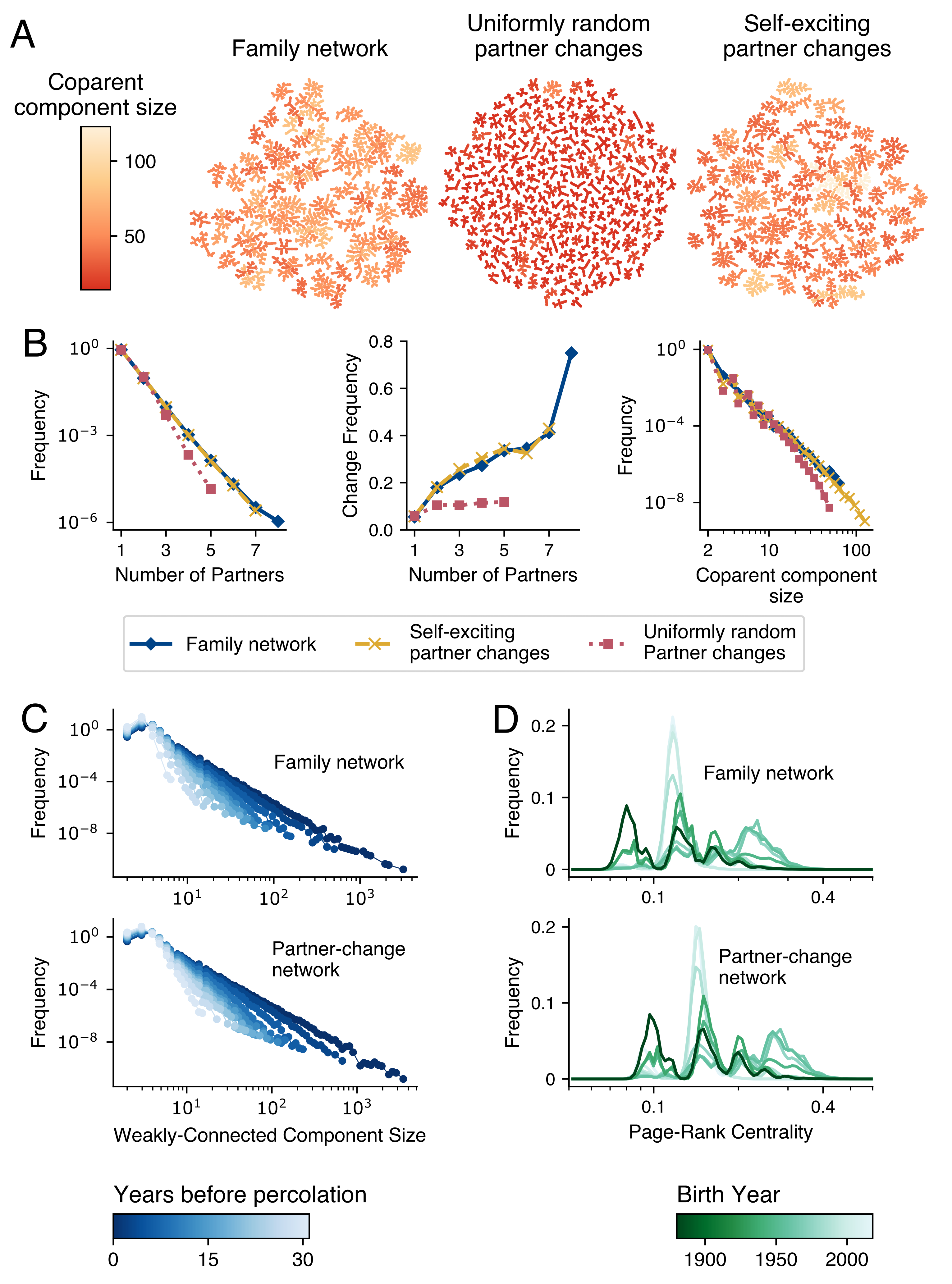}
    \caption{
    Partner-change behavior is heterogeneous and self-exciting.
    (A) The surrogate family-network model does not explain the sizes of coparent components in the Danish family network.
    (B) Partner-count distributions, partner-change frequencies stratified by prior-partner count, and connected coparent component size distributions of the Danish family network, networks generated assuming uniform random partner changes (surrogate family-network model), and assuming self-exciting partner changes (partner-change model).
    Modeling the observed self-exciting partner changes results in surrogate networks that better resemble the observed partner-counts and coparent component-size distributions as compared with assuming uniform random partner changes.
    By modeling the micro-structure of partner-change behavior, the resulting networks have strikingly similar dependencies between growth and structure as the empirical family network, exemplified in: 
    (C) weakly-connected component size distributions over time before the onset of percolation, and
    (D) birth-year stratified PageRank-centrality score distributions for a self-exciting network and the Danish family network.
    }
    \label{fig:self-exciting}
\end{figure}

A key metric that is not captured by the simple model is the size of components in the coparent network.
The Danish family network has a largest connected coparent component size of $82$, whereas the connected coparent components in the surrogate family networks are substantially smaller, with maximum component size $46\pm 6.24$.
This is visually very clearly expressed in Fig.\,\ref{fig:self-exciting}\,A, where a stark difference between the Danish family network and the surrogate family-network model (uniform random partner changes) is clearly visible.

The difference suggests that partner-change behavior in the empirical data is not homogeneous.
In fact, we find that partner-change behavior is heterogeneous in a systematic way, with partner-change frequencies increasing with previous-partner counts (Fig.\,\ref{fig:self-exciting}\,B).

Specifically, we find that previous-partner count is a significant predictor of partner change ($\alpha <0.01$) even when accounting for age, number of children with current partner, years since first child with current partner, and gender (see Appendix \ref{app:partner-change_regression} for details).
That partner-change behavior is self-exciting in this way is a novel finding with possible implications for the study of family patterns and genetics.

While novel in the family-network literature, self-exciting link-formation mechanisms, such as the preferential-attachment mechanism~\cite{Barabasi1999}, have been extensively studied in network science~\cite{newman2018networks} and are known to be important drivers of network structure in a range of other contexts~\cite{JeongLethality2001, AlbertError2000, BoccalettiStructure2006}. 

\textbf{Modeling self-exciting partner changes}.
To incorporate the observed self-exciting partner-change behavior in the surrogate family model, we stratify the yearly partner-change counts by individuals' number of prior partners, ensuring the correct number of transitions between partner-count groups over time (see Appendix \ref{app:self-exciting} for details), and refer to the resulting model and its generated networks as the \textit{partner-change model} and \textit{partner-change networks} respectively.
The partner-change model accurately captures the partner-count distribution, the dependency between prior-partner counts and partner-change frequency, and the coparent-component size distribution (Fig.\,\ref{fig:self-exciting}\,B) of the Danish family network.

Accounting for the yearly partner-change level and the self-exciting partner-change behavior, the partner-change model accurately captures additional complex attributes of the family network growth process, such as the dynamics of the component size distribution over time before the percolation (Fig.\,\ref{fig:self-exciting}\,C) and the PageRank centrality distribution of individuals in different birth-year cohorts (Fig.\,\ref{fig:self-exciting}\,D).

%%%%%%%%%%%%%%%%%%%%%%%%%%%%%%%%%%%%%%%%%%%%%%%%%%%%%%%%%%%%%%%%%%%%%%%%%%%%%%%%%%%%%%%%%%%%%%%%%%%%%
\subsection*{The effect of homophily}
%%%%%%%%%%%%%%%%%%%%%%%%%%%%%%%%%%%%%%%%%%%%%%%%%%%%%%%%%%%%%%%%%%%%%%%%%%%%%%%%%%%%%%%%%%%%%%%%%%%%%

Having mapped out the effects of partner-change behavior on the emergent large-scale structure of family networks, we can now meaningfully use our models to estimate the effect of partner-choices on the large-scale structure of family networks.

\begin{figure}[H]
    \centering
   \includegraphics[width=0.8\linewidth]{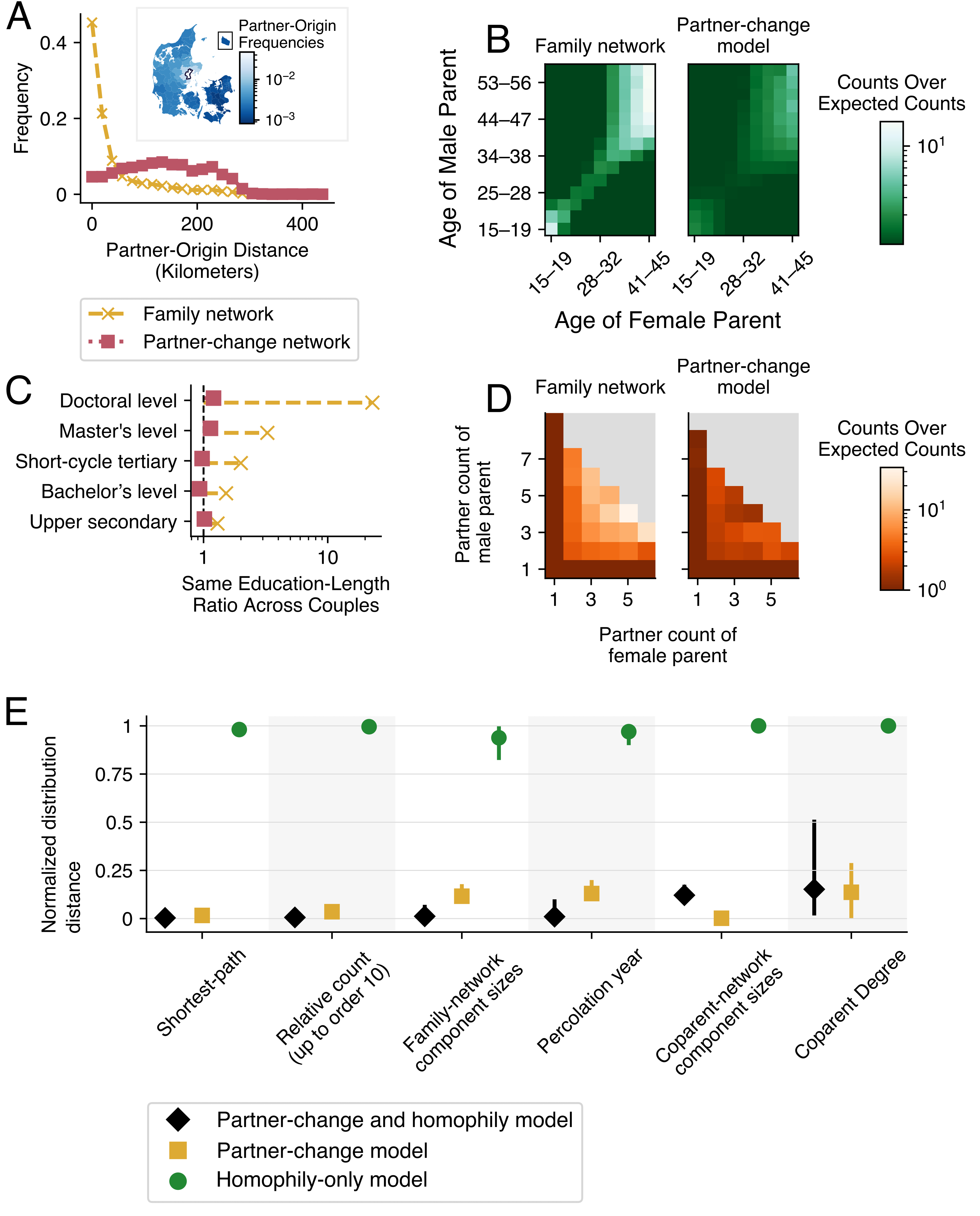}
    \caption{
        Homophily has limited effect on the emergent large-scale network structure.
        (A.1) Distance distribution between geographic origins of couples.
        (A.2) Empirical partner-origin frequencies conditioned on ego origin, normalized by population density, revealing that partnering probabilities are affected by natural barriers such as large bodies of water.
        (B) Age-combination counts across couples, normalized by the counts expected at random.
        (C) The frequency with which both partners in couples have the same education length normalized by the expected frequency, revealing clear homophily in the observed family network.
        (D) Prior-partner counts across couples normalized by expected counts.
        (E) Mean normalized distances between structural summary-statistic distributions of generated and empirical network measured by Earth mover's distance~\cite{rubner2000earth} or root-mean-square error (see Appendix \ref{app:model_accuracy} for details and Appendix \ref{app:homophily_unnorm} for the unnormalized results).
        Vertical lines show the span between minimum and maximum distance observed across 10 independent simulations.
        Incorporation of both the observed partner-choice biases and partner-change behavior in our model results in surrogate networks that only resemble the Danish family network slightly more than we achieve when only accounting for partner-change behavior.
       Failing to account for partner-change behavior result in networks structurally very dissimilar from the Danish family network, even when accounting for partner-choice homophily.
    }
    \label{fig:homophily}
\end{figure}

\textbf{Homophily is important in empirical pairwise partner choices}.
Using the Danish registries, we can retrieve socioeconomic information for both parents in each couple at the time of their first child and find that couples of the Danish family network display homophily with respect to geography, age, length of education, and prior-partner counts.
Birthplace inferred from residential registry data on municipality level (see Appendix \ref{app:homophily_model} for details) reveals that individuals which pair up are much more likely to have grown up in close proximity to each other than couples matched at random (Fig.\,\ref{fig:homophily}\,A).
The partner-origin frequencies conditioned on one part of the couple having grown up in a specified municipality, normalized by population density, reveals that partnering probabilities are also affected by natural barriers, such as large bodies of water (Fig.\,\ref{fig:homophily}\,A (inset), with the origin municipality in question highlighted with a black boundary line).
As expected, measuring the birthplace-distance distribution across couples in the partner-change network shows little geographical homophily, as the matching patterns are simply driven by by population density.

Computing the age combination counts across couples normalized by the expected counts under uniform random matching reveals that individuals which pair up are likely to be of similar age, with a tendency of the male partner to be slightly older, especially for among the oldest individuals (Fig. \ref{fig:homophily}\,B).
We observe a similar but less pronounced pattern of age combinations for couples in the partner-change network.
This is because in the model, coupled individuals are still more likely than chance to have similar ages in the generated network because age correlates with out-degrees, which we have to account for to form meaningful couples (see Appendix \ref{app:surrogate_family} for details).
However, the age combinations of couples in the Danish family network are again more homophily-driven than those of the partner-change network, confirming that age homophily is more pronounced than we would expect by chance in this network, even when considering the inherent bounds on the set of possible edge-configurations.

Computing the frequency with which both partners in a couple have the same education length (normalized by the expected frequency under uniform random matching) reveals clear homophily with respect to education length in the Danish family network (Fig. \ref{fig:homophily}\,C).
Partner-choice homophily with respect to education are absent in the partner-change network.

Finally, computing the prior-partner counts across couples (normalized by expected counts under uniform random matching) reveals that individuals with one or more prior partners are overly likely to pair up into a couple (Fig.\,\ref{fig:homophily}\,D).
Whereas homophily with respect to age, education length, and geographic origin have been treated extensively in the literature~\cite{Horwitz2023}, such partner count homophily has received little attention, and only ever as the binary distinction of having a prior partner or not~\cite{McDermott2013, Thomson2020Multi-Partner}.
We find, however, that a binary variable is too simple to fully capture partner count homophily, as real-world couples are increasingly more likely (compared to random) when the partner counts of both members increase.
We do observe a similar but less pronounced pattern of prior-partner count combinations in the surrogate family network, which emerges because we need to control for out-degree to form meaningful couples (see Appendix \ref{app:surrogate_family} for details).
However, the partner-count homophily that we observe for the Danish family network is noticeably more pronounced than for the partner-change network.

All of these these partner-choice biases segment the family network locally, resulting in individuals being more likely than random to connect with others of similar sociodemographic background, consistent with the literature on partner choice ~\cite{Horwitz2023, Derraik2015, Kaplanis2018, Lin2010}. 

\textbf{Homophily and large-scale network structure}.
To understand the effect of homophily on the emergent large-scale structure, we compare the accuracy with which the following three network models are able to capture different structural attributes of the Danish family network:
\begin{enumerate}
\item the partner-change model where we correct for couple-formation biases so as to match the empirical couple attribute concurrence distributions, which we denote the \textit{partner-change and homophily model};
\item the \textit{partner-change model} with uniform random couple formation;
\item the surrogate family-network model correcting for couple-formation biases but assuming no partner changes, which we denote the \textit{homophily-only model}.
\end{enumerate}
When accounting for partner-choice homophily, we parametrize trait heritability of geographic origin and educational level from parents to their children using a simple homogeneous mechanism where children inherit the features of a random parent (see Appendix \ref{app:homophily_model} for details).

For each of these three models, we generate 10 independent networks, compute distributions of a range of network-science metrics, and compare how well these distributions match what we find in the Danish family network.
We measure the differences between distributions using the Earth mover's distance~\cite{rubner2000earth} or root-mean-square error depending on the statistic in question (see Appendix \ref{app:model_accuracy} for details).
For each statistic considered, we normalize the scores to the interval $[0, \, 1]$ across models, enabling comparison across statistics (see Appendix \ref{app:homophily_unnorm} for the unnormalized results).
Thus, if two networks are similar, their distance will close to zero, and if the networks are very different their distance will be close to one.

We find that the partner-change and homophily model only matches the empirical data slightly better than the partner-change model in terms of undirected shortest-path distribution (shortest-path), average relative count up to order 10, family-network weakly-connected component size distribution at percolation, and the onset of percolation (Fig.\,\ref{fig:homophily}\,E).
Further, for capturing the connected coparent component size distribution and partner-count distribution, the partner-change model performs equally well or better than the partner-change and homophily model.
The homophily-only model is much worse at modeling all large-scale networks traits as compared with the partner-change model with and without homophily.
This analysis implies that accounting for partner-choice homophily only provides slight improvements in the realism of the generated network's large-scale traits as relative to the difference it makes to accounting for partner-change behavior.

We find that accounting for partner-choice biases slightly decreases the average shortest-undirected-path length from $40.97\pm 0.10$ to $40.18\pm 0.13$ and slightly accelerates the onset of percolation from occurring after $35$ years to instead occur after $34$ years (no variance across 10 independent simulations).
This may seem surprising, as homophily is generally thought to increase network distances~\cite{McPherson2001Birds} and delay the emergence of large components~\cite{Pham2021Balance, Murase2019, McPherson2001Birds}.
However, in the case of our experiments, homophily has the opposite effect because partner-count homophily brings highly connected individuals closer together in a similar manner to how degree assortativity affects other social networks~\cite{Noh2007Percolation}.

%%%%%%%%%%%%%%%%%%%%%%%%%%%%%%%%%%%%%%%%%%%%%%%%%%%%%%%%%%%%%%%%%%%%%%%%%%%%%%%%%%%%%%%%%%%%%%%%%%%%%
\section*{Discussion}
%%%%%%%%%%%%%%%%%%%%%%%%%%%%%%%%%%%%%%%%%%%%%%%%%%%%%%%%%%%%%%%%%%%%%%%%%%%%%%%%%%%%%%%%%%%%%%%%%%%%%

Homophily is often argued to play a large role in shaping social systems~\cite{branoulle2012LongRun, UganderAnatomy2011, BoccalettiStructure2006, McPherson2001Birds}.
Here, in the case of family networks, we observe a surprisingly limited effect of partner-choice homophily on the large-scale structural properties of family networks.
Instead, we find that partner-change behavior is a central factor, with partner-changing individuals connecting multiple otherwise distant families, affecting the emerging structure in a similar way as shortcuts do in the Watts-Strogatz model~\cite{Watts1998}.
Investigating which individuals change partner at what time reveals that partner-change behavior is a heterogeneous phenomenon that follows a self-exciting tendency where the frequency of changing partner increases with the cumulative number of prior partners.
Such self-exciting linking tendencies have been found in other network systems~\cite{Barabasi1999} and have important implications for the large-scale structure of networks~\cite{JeongLethality2001, AlbertError2000, BoccalettiStructure2006}.
We find that self-exciting partner-change behavior has clear structural consequences for family networks by creating large connected components in the resulting coparent networks.
Correcting for the absolute number of partner changes and their self-exciting tendencies, we are able to model the emergence of many large-scale structural traits of family network, such as the shortest path-length distribution, the percolation behavior, coparent component size distribution, component-size distribution over time, and relation between birth-year and network centrality.
Including homophily in partner-choices only marginally improves our ability to capture empirical large-scale structure when also accounting for partner-change behavior.
Further, failing to include partner-change behavior leads to unrealistic model networks, even when correcting for homophily.
In this sense we argue that that partner-change behavior is a more fundamental mechanism than partner-choice homophily when explaining the large-scale structure of family networks.

A limitation of our study is that we do not account for the full heterogeneity of socioeconomic-trait heritability, but instead assume simple rules of homogeneous heritability from parents to their children. 
However, heritability of education attainment, fertility-behavioral traits, and economic wealth are known to be complex processes, and we believe that modeling these heterogeneities could be an important next step in understanding the structure of family networks.

A further limitation is that the family network of Denmark is not necessarily representative of family networks as a whole, and we expect that some of the network properties discussed here may reflect particular cultural and societal characteristics of our population, and that these may not generalize to different times and societies.
However, networks with majority of single-partner individuals and a minority of partner-changing individuals is common across many countries~\cite{Thomson2020Multi-Partner}.
Therefore, we believe that the structural consequences of partner-change behavior are likely comparable across many family networks.

In summary, our work provides a quantitative investigation of the rich individual-level behavior underlying the large-scale structure of family networks, revealing that while edge-formation is indeed homophily-driven as argued in the literature, mechanisms from network science (small world effect and preferential attachment) have more profound structural consequences.

\printbibliography

%%%%%%%%%%%%%%%%%%%%%%%%%%%%%%%%%%%%%%%%%%%%%%%%%%%%%%%%%%%%%%%%%%%%%%%%%%%%%%%%%%%%%%%%%%%%%%%%%%%%%
%%%%%%%%%%%% APPENDECES %%%%%%%%%%%%
%%%%%%%%%%%%%%%%%%%%%%%%%%%%%%%%%%%%%%%%%%%%%%%%%%%%%%%%%%%%%%%%%%%%%%%%%%%%%%%%%%%%%%%%%%%%%%%%%%%%%
\appendix
\section*{Anonymity in Figures and Statistics}
All figures and summary statistics have been made in accordance with the anonymity requirement of Statistics Denmark to ensure that no individual is identifiable beyond membership of groups of 3 or more individuals.

%%%%%%%%%%%%%%%%%%%%%%%%%%%%%%%%%%%%%%%%%%%%%%%%%%%%%%%%%%%%%%%%%%%%%%%%%%%%%%%%%%%%%%%%%%%%%%%%%%%%%
\section{The Danish Family Network}\label{app:danish_family_network}
The parent-child relations from which the Danish family network is constructed are all from the national Danish population registries, which cover all individuals living in Denmark from 1968 until the present, augmented with archival data on marriages and births from the Danish national archives to reconstruct as many relations between parents and children alive in 1968, resulting in highly accurate family relation data from 1953 and onwards~\cite{Nielsen2024NationScale}.
To reduce bias caused by immigration, we restrict our analysis to the largest weakly-connected component.
The resulting network spans 65 years of observations and contains $\sim\,$6 million nodes and $\sim\,$8 million edges.

%%%%%%%%%%%%%%%%%%%%%%%%%%%%%%%%%%%%%%%%%%%%%%%%%%%%%%%%%%%%%%%%%%%%%%%%%%%%%%%%%%%%%%%%%%%%%%%%%%%%%
\section{Measuring Partner Changes}\label{app:measuring_partner_changes}
We use the following procedure to quantify the number of partner-changes per year: first, we identify the earliest recorded instance of each coparent pair and record the year of that observation, \( t' \).
Second, we track each individual’s partner changes, counting the number of such changes occurring in each year \( t \), denoted as \( N^t_{\text{change}} \).  
Importantly, \( N^t_{\text{change}} \) does not measure the number of new couples forming in year \( t \) but rather the number of partnerships that dissolve.
Therefore, if a couple is formed where both individuals' immediate prior child were with other partners, the formation of this couple counts as two changes of partners.
We aggregate partner changes on a yearly basis, as the vast majority of individuals have only one partner per year.

The number of partner-change counts that occur per year divided by the maximum number of possible changes, computed as the number of active parents that have a child with a partner in the given year, gives the partner-change level reported in the main text.

Our empirical yearly partner-change levels are affected by the left-censoring of our observations.
Therefore, we are unable to correctly count the number of partner changes that occur at early years: when observing an early parent-child link, we do not have sufficient information about the parent's prior relations to evaluate whether the link signifies a change of partner 
However, measured partner-change levels stabilize after 1984, from which point we believe left-censoring to have little effect.
However, as we only attempt to account for observed partner-change counts and not reason about their absolute values, this bias is not a major issue for our study.

%%%%%%%%%%%%%%%%%%%%%%%%%%%%%%%%%%%%%%%%%%%%%%%%%%%%%%%%%%%%%%%%%%%%%%%%%%%%%%%%%%%%%%%%%%%%%%%%%%%%%
\section{Surrogate family-network model}\label{app:surrogate_family}
\begin{figure}[H]
    \begin{center}
        \includegraphics[width=0.75\textwidth]{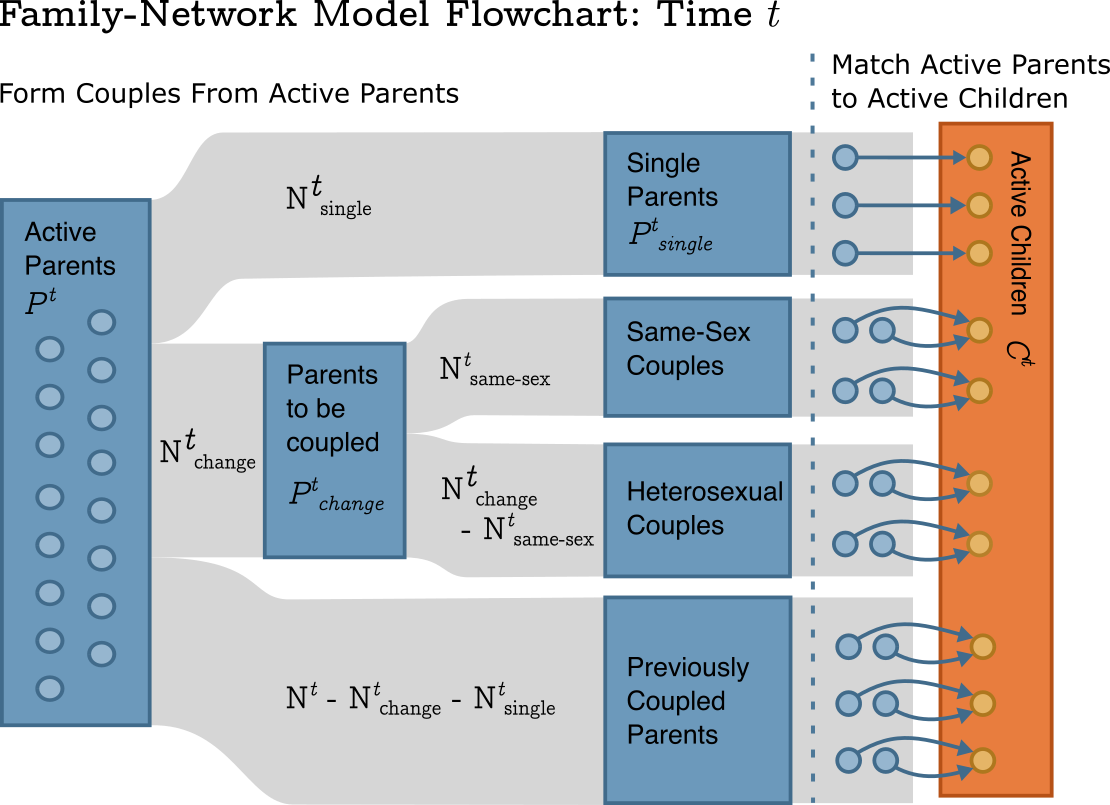}
    \end{center}
    \caption{
        Volume-chart for the surrogate family-network model.
    }\label{fig:family_model_volumechart}
\end{figure}

Here, we give full details for the surrogate family-network model as introduced in Sec. \ref{sec:partner_change_counts} of the main text.
Among the details of the surrogate family-network model that were left out of the main text to ensure clarity, the most important are the stratification of nodes by sex, the saturation of nodes when they reach their desired out-degree, matching parents into couples based on how many children they still want to have, the stratification of couples by same-sex and heterosexual relationships, and the sequential child assignment procedure to ensure as large a parent-child matching as possible.
We summarize these in Table \ref{tab:family_model_notation} and visualize the sampling and matching procedure in Figure \ref{fig:family_model_volumechart}.

Let there be given a directed family network with a nodeset of individuals $V$ and an edge set $E$ of time-stamped and directed edges that each go from a parent to one of their children.
We choose to work with yearly aggregations of timestamps.
With a slight abuse of notation we let $T:E\rightarrow\mathbb{N}$ be the mapping from edges in the network to its discrete timestamp.
Let each individual be assigned a sex and let $S:V!\to!{\mathrm{Male},\mathrm{Female}}$ be the mapping from nodes to their sex.
If a node has no registered sex, we assign it a sex (male or female) uniformly at random.
For the Danish family network, more than $99\%$ of nodes have a registered sex.

The family-network model generates a new directed acyclic network with the same nodeset of individuals as the family network and a new edge set of time-stamped edges $\hat{E}$ such that for any node $i$, number of in- and out-edges and their timestamps are as similar between $E$ and $\hat{E}$ as possible.
Generating accurate in- and out-degree sequences and the time-stamps of edges accounts to controlling for individual-level fertility behavior which is otherwise difficult to account for~\cite{RugglesConfessions1993}.

We build surrogate family networks iteratively, adding nodes in the order of their birth year and connecting them to a set of parents where at least one parent has an out edge in the given year.

For a given year $t$, and for the subset of edges that have timestamp $t$, we refer to the set of nodes that are the targets of at least one edge in this subset as \textit{active child nodes} and the set of nodes that are sources of at least one such edge as the \textit{active parent nodes}.

If two nodes have a child together, both being sources of edges that target the same node and have identical time stamps, then they constitute a \textit{couple}.
Nodes in a couple are \textit{partners} of one another.
A couple lasts until either parent has a child with a new partner.

For a given year $t$, we say that a node is \textit{saturated} if it has been assigned enough children to match its out-degree in the observed family network.
We denote the set of \textit{unsaturated active parents} at year $t$, to be the subset of active parents where neither the parent nor their partner has been saturated.
We discuss the effects of saturating parents this way in Appendix \ref{app:validate_model}.

We denote the set of active children and the set of unsaturated active parents at year $t$ as $C^t$ and $P^t$ respectively.

To generate the synthetic family network, for each year $t\in T$, we start by splitting the set of unsaturated active parents into three groups: parents that have a child without a partner, parents that in a couple and stay with their partner, and a group of parents that either don't have a partner or change partner at year $t$.

To this end, we first count the number of active parents in the observed network that are not part of a couple at year $t$ stratified by gender.
We denote the empirical number of men and women that have a registered child without any partner at year $t$ as $N^t_{\mathrm{single male}}$ and $N^t_{\mathrm{single female}}$ respectively.
At year $t$, we uniform randomly sample $N^t_{\mathrm{single male}}$ and $N^t_{\mathrm{single female}}$ male and female nodes from the set of unsaturated active parents that have not been assigned any children at time $t'<t$. 
We designate these to be single parents at year $t$ in the generated network and denote this set $P^t_{\mathrm{single}}$.
We denote the set of active parents that must be coupled $P^t_{\mathrm{couple}} \,=\, P^t \, - \, P^t_{\mathrm{single}}$.

For the Danish family network, we observe that most single parents are also \textit{first-time parents}, that is, active parents that have not been active at any prior time.
Denote this set by $P^t_{\mathrm{first}}$.
Therefore, we choose to sample single parents exclusively from $P^t_{\mathrm{first}}$.
We have not investigated the effects of this modeling decision.

We sample the parents that will change partners among the coupled and unsaturated active parents.
To ensure the correct number of partner changes, we measure the \textit{partner-change count} at time $t$: the number of active parents with distinct partners at time $t$ and $t-1$ in the observed network (see \ref{app:measuring_partner_changes} for further details).
We denote the partner-change count as $N^t_{\mathrm{change}}$ and the set of parents that change partner as $P^t_{\mathrm{change}}$.

If an active parent was previously in a couple, but was left by its partner, then coupling this node will count as a partner-change.
To account for these inevitable partner-changers, at year $t$, we add all nodes that had a partner in year $t'<t$ but no partner in year $t$ to $P^t_{\mathrm{change}}$ and subtract the number of such nodes from $N^t_{\mathrm{change}}$.
    
Among the coupled and unsaturated active parents at time $t$ that are not inevitable partner-changes, we sample $N^t_{\mathrm{change}}$ nodes using the following procedure: Sample a single node and add it to $P^t_{\mathrm{change}}$, then check if its current partner is also an active parent at time $t$, if so we add the partner to $P^t_{\mathrm{change}}$. 
We repeat this procedure until $|P^t_{\mathrm{change}}| \geq N^t_{\mathrm{change}}$ or until we have no more active coupled parent to sample from.
Finally, we add all active parent from the set of first-time non-single active parents $P^t_{\mathrm{first}} \, \backslash \, P^t_{\mathrm{single}}$ to $P^t_{\mathrm{change}}$.

We now match the active-parent nodes $P^t_{\mathrm{change}}$ into pairs.
The active-child nodes will later be distributed among these pairs.
When forming pairs, to ensure the right number of same-sex pairs at year $t$, we count the number of same-sex couples at denote them $N^t_{\mathrm{male-male}}$ and $N^t_{\mathrm{female-female}}$.
We then sample $N^t_{\mathrm{male-male}}$ and $N^t_{\mathrm{female-female}}$ uniform random unsaturated active parents from $P^t_{\mathrm{change}}$, which are matched to form male-male and female-female pairs. 
The unsaturated active parents in $P^t_{\mathrm{change}}$ that have not been assigned into same-sex pairs are matched to form heterosexual pairs.

To form couples out of a set of unsaturated active parents, we split the set into \textit{egos} and \textit{partners}.
For heterosexual pairs, we select all the male nodes to be egos and female nodes to be partners.
Each ego is then assigned a uniform random partner until there are no unmatched egos or partners.

Further, to better match the observed out-degree distribution, we match all pairs of unsaturated active parents to have the same number of remaining-activation counts: the empirical out-degree of a node subtracted the number of children assigned to it up to time $t-1$.
Sex combinations or group sizes can prohibit us from performing perfect matches within remaining-activation count groups, leaving us with a set of uncoupled individuals.
We attempt to match these into heterosexual couples uniformly at random.
Any node not matched by this procedure is treated as a single parent and added to $P^t_{\mathrm{single}}$.
This matching scheme is important for generating realistic out-degree distributions, and we discuss its implications in Appendix \ref{app:out-degree_seq}.

Finally, we match active children with unsaturated active parents at time $t$.
We perform this matching in an out-degree biased manner, using the out-degree of all unsaturated active parents and children in the observed family network as well as the empirical out-degree cooccurrence distribution of year $t$.
For each active child $c$ in $C^t$, we sequentially sample parents that have not already been assigned an edge at time $t$ according to the child's out-degree, the remaining unsaturated active parents out-degree, and the out-degree co-occurrence distribution.
Let $i$ be a sampled parent.
We add the directed edges $(i,\, c)$ to the edge set of the generated network $\hat{E}$.
If $i$ has a partner $j$ at time $t$, then we also add $(j,\, c)$ to $\hat{E}$.
We sample in this way to maximize the number of parents covered by the matching procedure.
If all unsaturated active parents have been assigned at least one edge at time $t$, but not all children have been assigned a set of parents, then we sample additional parents according to the out-degree co-occurrence distribution, adding directed edges from the sampled parent (and their current partner if they have one) to the child in question.
All edges assigned this way are given year-aggregated timestamp $t$.

We repeat this procedure for all $t\in T$ in ascending order. 
The node set of individuals and the set of directed and time-stamped edges $\hat{E}$ constitute the generated network.

\begin{table*}[h!]
    \centering
    \renewcommand{\arraystretch}{1.05}
    \begin{tabularx}{\textwidth}{l X}
        \textbf{Symbol} & \textbf{Meaning} \\
        \hline
        $V$ & Node set (individuals). \\
        $E$ & Edge set of directed, time-stamped parent\,$\to$\,child links. \\
        $T$ & Set of discrete years observed. \\
        $\tau:E\!\to\!\mathbb N$ & Timestamp function ($\tau(e)=t$ assigns edge $e$ to year $t$). \\
        $S:V\!\to\!\{\mathrm{M},\mathrm{F}\}$ & Sex of each node. \\[2pt]
        $C^t$ & Active children at year $t$ (targets of edges with $\tau(e)=t$). \\
        $P^t$ & Active parents at year $t$ (sources of edges with $\tau(e)=t$). \\
        $P^t_{\mathrm{single}}$ & Unsaturated active parents having a child in $t$ without a partner. \\
        $P^t_{\mathrm{first}}$ & Active parents in $t$ with no earlier activity ($t' < t$). \\
        $P^t_{\mathrm{change}}$ & Active parents that change partner in year $t$. \\[2pt]
        $N^t_{\mathrm{single},\,\mathrm{M/F}}$ & Number of single fathers / mothers in year $t$. \\
        $N^t_{\mathrm{change}}$ & Partner-change count in year $t$. \\
        $N^t_{\mathrm{MM}}$, $N^t_{\mathrm{FF}}$ & Numbers of male–male / female–female couples in $t$. \\
        \hline
    \end{tabularx}
    \caption{Notation used in the surrogate family-network model (Sec.~\ref{sec:partner_change_counts}).}
    \label{tab:family_model_notation}
\end{table*}

%%%%%%%%%%%%%%%%%%%%%%%%%%%%%%%%%%%%%%%%%%%%%%%%%%%%%%%%%%%%%%%%%%%%%%%%%%%%%%%%%%%%%%%%%%%%%%%%%%%%%
\section{Validating the surrogate family-network model}\label{app:validate_model}

The aim of the surrogate family-network model is to generate networks that account for individual-level fertility behavior.
Here, we validate the models ability to account for the observed out-degree sequence and the out-degree correlation between parents and their children, as we found this to be the main challenges of accurately mimicking real fertility behavior.
We discuss the following model features: \textit{parent saturation}, \textit{remaining-activation-counts grouped coupling of parents}, and \textit{out-degree biased parent-child matching}.

\subsection{Aligning generated and empirical out-degree sequences}\label{app:out-degree_seq}
There are two reasons why the surrogate family-network model without out-degree restrictions and remaining-activation counts grouped coupling is misaligned with the observed network.
Firstly, individuals in couples can be active in non-identical sets of years.
When this occurs, couples can be assigned more children than either parent has in the observed network, because, when one parent is active, the other parent is also eligible for being assigned a child.
This effect tends to increase the out-degree of nodes.
Therefore, we refer to it as \textit{out-degree inflation}.

Second, couples are not guaranteed to be assigned a child when one or more of its members are active.
This occurs because the set of active parents can have partners that are not active. 
In the observed network, the majority of nodes have two parents.
Therefore, for each year, the number of active parents in the surrogate family-network model is approximately twice that of active children.
However, when active parents have inactive partners, the ``effective size'' of the active parent set can be over twice that of the active child set, making it impossible to assign each couple a child for the given year.
Empirically, we observe that the set of active non-saturated parents and their partners is $20\%$ too large to assign each couple a child.
Because this effect tends to decrease the degree of nodes, we refer to it as \textit{out-degree deflation}.
            
We start by investigating the effect of parent saturation and remaining-activation counts grouped coupling on the generated out-degree distribution.
To this end we perform an ablation study of the surrogate family-network model described in Appendix \ref{app:surrogate_family}, where we alter the saturation criteria and coupling matching procedure.
    
First, we generate a network using the family-network model disregarding parents saturation and where couples are matched uniformly at random without remaining-activation counts grouping.
That is, for each year, all active parents are eligible to receive a child and there is no enforced alignment on the out-degree of partners.
We denote this network the \textit{naive model network}

Second, we generate a network using the surrogate family-network model where nodes only saturate when they themselves reach their desired out-degree (number of assigned children), irrespective of whether their partner has saturated, and couples are matched uniformly at random without remaining-activation counts grouping.
We denote this network the \textit{ego-saturated model network}

Third, we generate a network using the surrogate family-network model with the correct couple-based saturation criteria but where couples are still matched uniformly at random without any remaining-activation counts grouping.
We denote this the \textit{couple-saturated model network}.

Fourth, we generate a network using the full family-network model, where we enforce both the couple-based saturation criterion and create couples within remaining-activation count groups.

\begin{figure}
    \centering
        \includegraphics[width=0.95\textwidth]{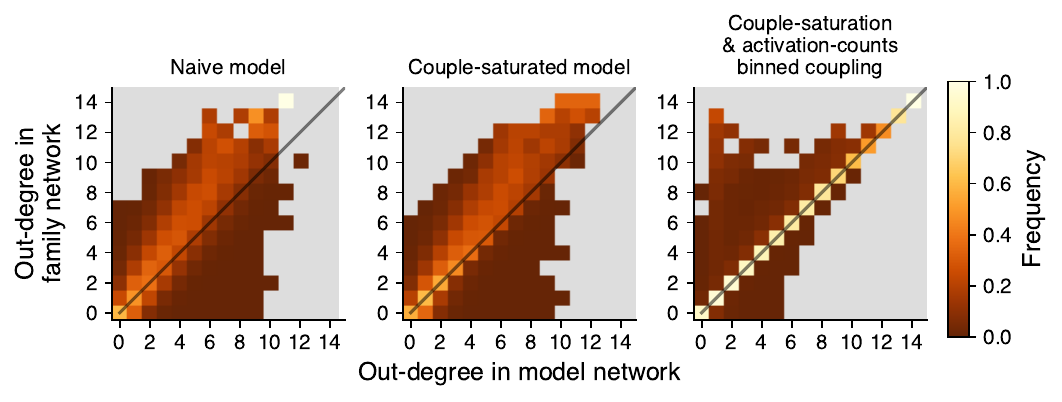}
    \caption{Comparison of the out-degree sequence of the Danish family network and networks generated using the surrogate family-network model with and without node saturation and remaining-activation matching.
        We see that node saturation and remaining-activation matching improve the ability of the surrogate family-model to generate networks that match the out-degree sequence of the original network.
    }\label{fig:outdegree_match}
\end{figure}

By plotting the out-degree of each node in the Danish family network and for each of the above networks, we can see the effects of our saturation criteria and remaining-activation counts grouped coupling.
The out-degrees for each node in the observed and naive model network reveals that without saturation and grouped coupling, the resulting out-degree sequence differs greatly from the observed one showcasing signs of both out-degree inflation and out-degree deflation, as we see in Figure \ref{fig:outdegree_match}).
Notably, we observe no clear diagonal line, revealing that the majority of nodes are assigned the wrong out-degree by the model.

The ego-saturated model network exhibits overall less out-degree inflation than the naive model network.
However, we do still observe inflation, notably among the low-degree nodes.
This inflation of small out-degrees occurs whenever a low-degree node and a high-degree node form a couple, as the number of children the couple is assigned will scale with the max of their out-degree.

The couple-saturated model network avoids much of this low out-degree inflation by removing parents from the active-parent set whenever they or their partner are saturated.
However, this creates out-degree deflation as couples of a low-degree node and a high-degree node stop having children when the low-degree node is saturated .
            
To alleviate this deflation, we restrict couples to form within remaining-activation count groups in the surrogate family-network model, centering the generated out-degrees on the observed values in Figure \ref{fig:outdegree_match}.
We use remaining-activation counts instead of out-degree when grouping nodes before matching couples to ensure that each member of every couple will create the same number of connections, even if one or both members have had children with prior partners.

We believe forming couples within remaining-activation counts groups to be sensible from a modeling perspective, because individuals that pair up into child-bearing couples are likely to align expectations about their desired number of children.

Based on this ablation study, we find that enforcing a couple-based saturation criteria and remaining-activation counts grouped coupling is adequate to control for the out-degree sequence when generating synthetic family networks.

\subsection{Correcting for out-degree correlation across parents and children}
Beyond generating realistic marginal out-degree distributions, we also want to generate realistic correlation of out-degrees between parents and their children.

\begin{figure}
    \centering
        \includegraphics[width=0.95\textwidth]{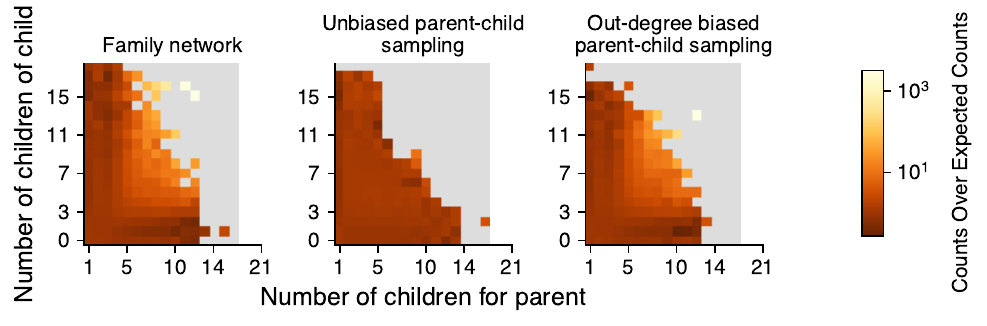}
    \caption{
        Out-degree co-occurrence of parents and their children for the Danish family network and surrogate family-model networks generated with and without out-degree biased parent-child assignment, normalized by the expected co-occurrence under uniform random parent-child matching.
        We see that nodes with parent-child edges are overly likely between nodes of high out-degree, and that biased out-degree parent-child assignment is necessary to accurately generate networks that respect this tendency.
    }\label{fig:outdegree_parent_child}
\end{figure}

Plotting the out-degree co-occurrence counts for parents and their children in Figure \ref{fig:outdegree_parent_child}, reveals that high out-degree parents are more likely to have high out-degree children that would be expected at random.
The out-degree co-occurrence counts of a family-model network with uniform parent-child sampling does not recreate this correlation pattern.

To enforce out-degree correlations correctly, we use the yearly parent-child out-degree co-occurrence distributions.
This accounts for shifts in out-degree co-occurrence patterns, such the tendency that children are much more likely to be leaf-nodes at times close to the right-censoring of our observation.

We believe directly biasing the parent-child matching by their out-degree in the observed model is meaningful.
The observed-network out-degree of a node can be thought of as a desired number of children for each individual.
Matching parents and children based on the observed-network out-degree can then be thought of as incorporating heritability of social norms about optimal family sizes.

We see no clear effect stratifying by gender of parent and child, so we work with a single out-degree co-occurrence distribution for each year.

In Figure \ref{fig:outdegree_parent_child}, we see that matching parents and children in accordance with the observed yearly out-degree co-occurrence distribution generates networks with realistic out-degree correlation patterns.

%%%%%%%%%%%%%%%%%%%%%%%%%%%%%%%%%%%%%%%%%%%%%%%%%%%%%%%%%%%%%%%%%%%%%%%%%%%%%%%%%%%%%%%%%%%%%%%%%%%%%
\section{Partner-Change Momentum Regression Analysis}\label{app:partner-change_regression}
Here we describe our regression analysis on individuals likelihood to change partner when having a new child.
We do this regression analysis to investigate whether the number of distinct prior partners has an effect on the likelihood of changing partner.

From the Danish family network, we find all individuals that have two or more children and register the times that these children are born (individuals with fewer registered children cannot inform us about the likelihood of changing partner).

For each individual and child-birth timestamp, we register the age; sex; the number of years since their most recent change of partner ($0$ if the individual has not changed partner previously); the number of years since their first child; and their number of distinct prior partners.
In addition we register if they have the current child with a new partner, encoded binary as $0:=\text{no new partner}$ and $1:=\text{new partner}$.
We restrict our analysis to those individuals where all covariates and partner-change outcomes are known, resulting in $4833922$ parent-time observations for $3146157$ unique individuals.

We fit a logistic regression to the likelihood of each individual to change partner as a function of the covariates listed above and present the results as a regression table in \ref{app:tab:partner-change_regression}.

We find that all the covariates are significant for $\alpha = 0.001$, and that the number of prior partners increases the likelihood of an individual to change partner.
\begin{table*}[htbp]
    \begin{tabular}{lcccccc}
        \hline\hline
        & Coefficient & std. error & z & $P>|z|$ & [0.025 & 0.975] \\
        \hline
        Intercept & -2.9681 & 0.016 & -187.451 & 	\textless 0.001 & -2.999 & -2.937\\
        Number of prior partners & 2.7333 & 0.010 & 265.495 & 	\textless 0.001 & 2.713 & 2.753 \\
        Number of Children & -1.2112 & 0.003 & -361.822 & 	\textless 0.001 & -1.218 & -1.205 \\
        Years since most recent change & 0.4059 & 0.002 & 265.569 & 	\textless 0.001 & 0.403 & 0.409 \\
        Sex ($0:=\text{Female}$, $1=\text{Male}$) & 0.2973 & 0.004 & 78.561 & 	\textless 0.001 & 0.290 & 0.305 \\
        Age & -0.1080 & 	\textless 0.001 & -235.529 & 	\textless 0.001 & -0.109 & -0.107 \\
        Years since first child & 0.0408 & 0.002 & 26.873 & 	\textless 0.001 & 0.038 & 0.044 \\
        \hline\hline
    \end{tabular}
    \caption{
    Regression table for logistic regression on partner change in the Danish family network. Individuals that have fewer than 2 children are not included in the regression analysis. All variables are found to be significant for $\alpha = 0.001$.
    }
    \label{app:tab:partner-change_regression}
\end{table*}

%%%%%%%%%%%%%%%%%%%%%%%%%%%%%%%%%%%%%%%%%%%%%%%%%%%%%%%%%%%%%%%%%%%%%%%%%%%%%%%%%%%%%%%%%%%%%%%%%%%%%
\section{The partner-change model}\label{app:self-exciting}
The surrogate–family model described in the main manuscript replicates many large-scale structural features of the Danish family network, but fails to capture the self-exciting nature of partner changes which we observe, leading to misspecified coparent network structure.  
Here, we give a concise description of the \textit{partner–change model}, which accounts for the heterogeneity of partner-change behavior.

The partner-change model is identical to the surrogate family model in all respects except one; we replace the uniform sampling of activate parents to change partner with a skewed sampling scheme that accounts for the self-exciting partner-change tendency that we observe in the Danish family network.

To account for the increasing frequency of partner-changes with prior-partner count, we stratify the yearly partner-change count along two axes.  
Firstly, we stratify by individuals prior–partner count $p$; sex $s$; and \textit{remaining-activation count} $r$, defined as the difference between the individual’s empirical out–degree and the number of children already assigned up to $t$.  
The inclusion of remaining-activation count $r$ is crucial because it forms an upper bound on the number of additional partner a node can have: an individual with only two remaining activations and only a single prior partner can never attain a total partner count of five, and ignoring this constraint would bias the high-$p$ states downward.

Using the Danish family network we compile, for every year, a three–way lookup
$$
\Delta_{t}(p,r,s)\in\mathbb{N}
\qquad(p=0,1,2,\dots;\;r=0,1,2,\dots;\;s\in\{\mathrm{M},\mathrm{F}\}),
$$
which records how many parents of sex $s$ move from state$(p,r)$ in year $t-1$ to $(p+1,r-1)$ in year $t$.  
We refer to $\Delta_{t}$ as the \emph{stratified change table}. 
Its entries provide the target numbers of partner changers that must be drawn from each \((p,r,s)\) stratum.

When generating the network, at year t \(t\), we identify all \emph{coupled, unsaturated, active} parents and denote these the \textit{eligible set}.  
Parents whose partner left them at some year $t'<t$ and who are now active for the first time since then are unavoidable changers (as introduced in Appendix \ref{app:surrogate_family}.
These are added to the set of partner-changers $P^t_{\mathrm{change}}$ and we deduct their counts from the appropriate $(p, r)$ table entries before sampling further partner changers.
If this subtraction empties a $(p, r)$ cell, the remainder is subtracted from the same cell of \(\Delta_{t+1}\).

From the remaining eligibility pool we uniform randomly sample partner changes changes from the eligible set from the set of non-empty cell of \(\Delta_t\) in a uniform random order and add these to $P^t_{\mathrm{change}}$ until the number of partner-changers succeeds the yearly partner-change count $N^t_{\mathrm{change}}$, passing any unsatisfied quota forward to the corresponding cell of \(\Delta_{t+1}\).
This memory mechanism guarantees that the multi-year \((p,r,s)\) totals of the simulation converge to their empirical counterparts.

Individuals with five or more partners are so rare that the stratified table becomes sparse and noisy, making the stratification scheme inefficient at correctly modeling their partner-change behavior.  
Instead, we assign a fixed partner–change probability of $P_{\text{tail}} = 0.10$, for all individuals with 5 or more prior-partners whenever these are active, a value tuned to stabilise the upper tail of the partner-count distribution without distorting lower counts.

\begin{figure}
    \begin{center}
        \includegraphics[width=1\textwidth]{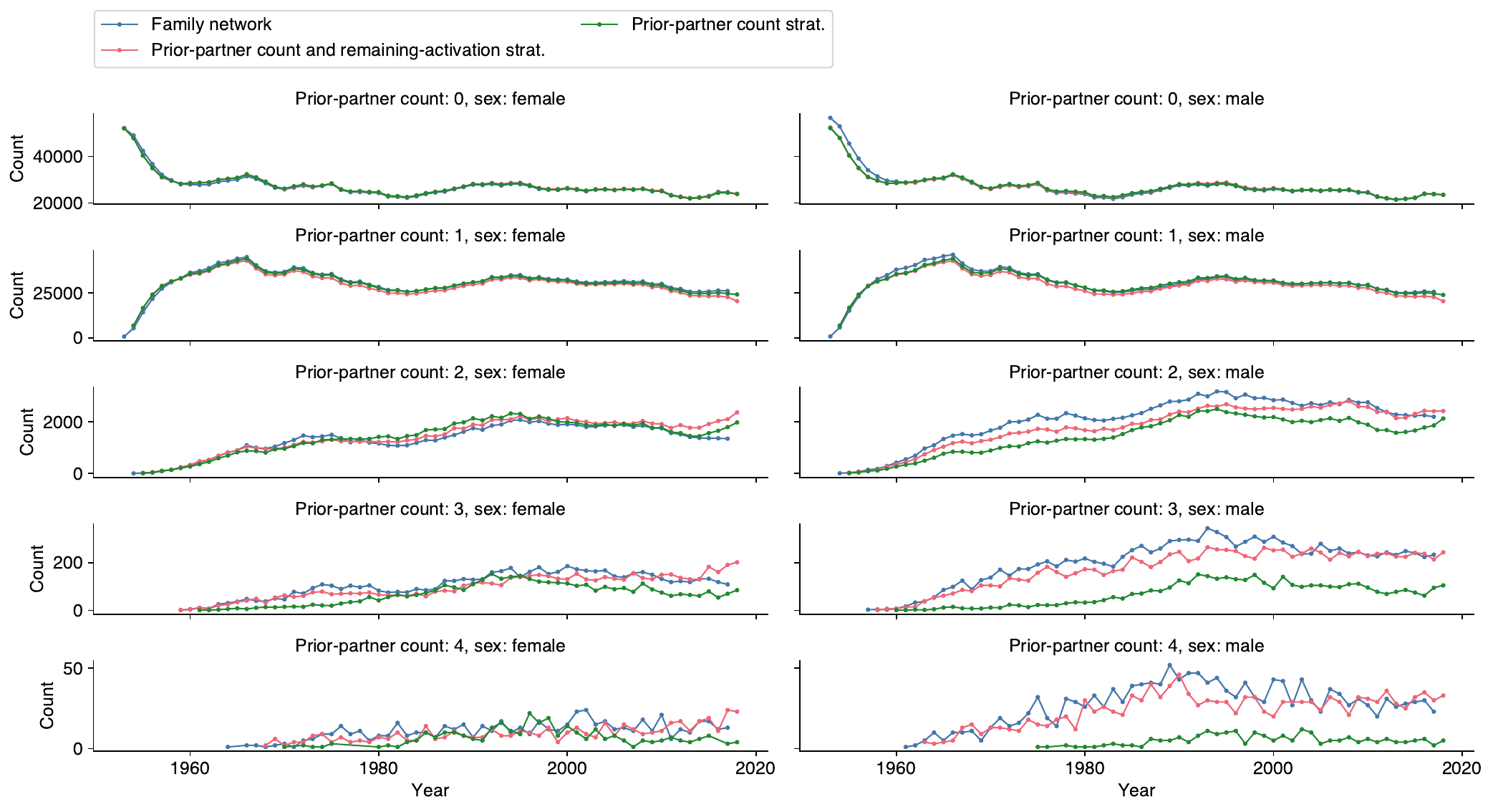}
    \end{center}
    \caption{
        The number of active parents over time in the Danish family network and for a family-network model where partner change counts are stratified by prior-partner counts, remaining activation, and sex, as opposed to just prior-partner counts and sex, visualized for partner-count groups ranging up to 4 partners.
        We see that to generate networks with number of high partner counts we need to also stratify the yearly partner-change counts by remaining activations.
    }\label{fig:partner_count_group_sizes}
\end{figure}

By comparing the prior-partner count group sizes over time for partner-count groups up to 4 in Figure \ref{fig:partner_count_group_sizes}, we can see the difference between stratifying partner-change groups by prior-partner count, sex, and remaining activation as opposed to only stratifying by prior-partner count.

%%%%%%%%%%%%%%%%%%%%%%%%%%%%%%%%%%%%%%%%%%%%%%%%%%%%%%%%%%%%%%%%%%%%%%%%%%%%%%%%%%%%%%%%%%%%%%%%%%%%%
\section{Correcting for partner-choice biases}\label{app:homophily_model}

The surrogate–family framework described above matches parents uniformly at random within \textit{remaining–activation} groups while ensuring correct yearly sex–combination counts.  
While sufficient for degree and timing statistics, this rule misrepresents the empirical partner-choice of the Danish network as illustrated in Figure \ref{fig:homophily} panel A in the main text.  
Partners tend to possess similar prior–partner counts, age, educational length, and childhood geography.  
To control for this sorting behavior, we introduce a two–level, biased matching mechanism that respects partner-choice homophily while preserving the bookkeeping logic of the baseline model.

We control for partner-count homophily differently than for sociodemographic homophily.
One potential issue of this methodology is that partner-count homophily is assumed to be independent of other types of homophily.
However, we have not investigated the effects of this decomposition of homophily.

\subsection*{Accounting for partner–count homophily}
Let $p\in\mathbb N$ denote the \textit{prior–partner count} of a parent and $r\in\mathbb N$ the \textit{remaining–activation count} (empirical out-degree minus births already assigned).  
For every calendar year $t$ we measure in the register the number of heterosexual couples whose member states are  
$$
\bigl(p_{\mathrm{M}},\,r_{\mathrm{M}}\bigr)\;\times\;\bigl(p_{\mathrm{F}},\,r_{\mathrm{F}}\bigr)\,,
$$
and store the result in the four–way lookup  
$$
\Gamma_t\!\bigl((p_{\mathrm{M}},r_{\mathrm{M}}),(p_{\mathrm{F}},r_{\mathrm{F}})\bigr)\in\mathbb N\,.
$$

We refer to $\Gamma_t$ as the \textit{partner–count} $\times$ \textit{remaining–activation target table}.  

For year $t$, we partition the coupled and uncoupled active parents into cells based on their partner count, remaining-activation count, and sex:
$$
(p,r,s)\quad s\in\{\mathrm{M},\mathrm{F}\}\,,
$$
exactly as already required by the partner–change model.  

We then perform matching using the following procedure:
We first sample the requested male–male and female–female couples strictly from the $(p=0)$ groups following the homophilious matching scheme described in Appendix subsection \ref{app:sociodemographic_matching} below.  
We then iterate over the non–zero entries of $\Gamma_t$ in random order, and for each requested cell, select the required number of men from $(p_{\mathrm{M}},r_{\mathrm{M}},\mathrm{M})$ and women from $(p_{\mathrm{F}},r_{\mathrm{F}},\mathrm{F})$.  
These are then matched in accordance with the observed sociodemographic matching distributions as outlined in Appendix subsection \ref{app:sociodemographic_matching} below.
If only $k<\Gamma_t$ couples can be formed for a given cell, the deficit is added to the corresponding cell of $\Gamma_{t+1}$.  
This memory mechanism guarantees that long-run frequencies of $\bigl(p,r\bigr) \times \bigl(p',r'\bigr)$ pairings match their empirical counterparts even if individual yearly targets cannot always be met.

Any active parents left unmatched after performing the homophilious matching procedure described above are coupled by the out-degree–binned matching scheme from the surrogate family-network model, which falls back to perform uniform random matching of male-female couples to achieve as large a matching as possible.  

\subsection*{Accounting for socioeconomic homophily}\label{app:sociodemographic_matching}
To perform sociodemographic homophilious matching, we first compute the observed partnering frequencies across age, computed as current year minus the year of the couple formation; educational length, measured as the highest ISCED–P level of a completed education before the date of couple formation; and geographic origin, measured as the municipality where the individual has been registered to live the longest before age 18.

National registries cover all completed educations from 1971 and onward, and residency from 1986 onward.
Therefore, full trait tuples are only observable from year $\approx2000$ and onward; the class–combination frequencies $\Psi$ necessarily reflect late-period assortativity.
Applying them retroactively to births in the 1950s and 1960s may therefore fail to reproduce the educational and geographic homophily in those early years.  

For every observed couple we discretize the triplet (age bin, origin code, education level) into a single categorical \textit{class}.  
Counting all unordered class pairs yields empirical frequencies  
$$
\Psi\bigl(c_{\mathrm{ego}},c_{\mathrm{partner}}\bigr)\;,
$$
which we store in a sparse dictionary with Laplace smoothing (default value $10^{-10}$).

During simulation the pool of \textit{eligible} egos and partners is first split by these classes.  
Couples are then sampled by a greedy Monte-Carlo procedure:
First, we draw an ego class at random; within it pick an ego $i$.  
We then sample a partner class $c$ with probability proportional to $\Psi(c_{\text{ego}},c)$.
We then sample a random partner $j$ from class $c$ and accept the pair if $j$ is not $i$’s current partner and $j \neq i$, removing $i$ and $j$ from the eligible pool until the set is empty.  
Any unmatched parents remaining after the homophily pass are coupled uniformly at random to form heterosexual couples.

Together, the partner–count $\times$ fertility–state table $\Gamma_t$ and the sociodemographic frequency table $\Psi$ steer couple formation toward the empirically observed heterogeneity without sacrificing the global balance constraints enforced by the surrogate–family backbone.

%%%%%%%%%%%%%%%%%%%%%%%%%%%%%%%%%%%%%%%%%%%%%%%%%%%%%%%%%%%%%%%%%%%%%%%%%%%%%%%%%%%%%%%%%%%%%%%%%%%%%
\section{Measuring network model accuracy}\label{app:model_accuracy}
Here we detail how we compare the similarity the generated networks and the Danish family networks that was presented in Figure \ref{fig:homophily} of the main manuscript.

For each network model in question (partner-change and homophily model, partner-change model, monogamous homophily model) we generated 10 independent networks.
For each of these generated networks and for the Danish family network we computed the following statistics: 1) the shorted undirected-path distribution as sampled from 2000 uniform random pairs for the given network; 2) the average relative count of relative-order ranging from 1 (number of parents and children) to 10 (number of relatives reachable by a path of length 10 and no shorter path); 3) parent-child network's weakly connected component size at the onset of percolation, measured as the year where the largest weakly connected component grew to make up $10^{-3}$ of the network; 4) the percolation year; 5) the coparent component size distribution; and 6) the partner-count degree distribution.

We then measured the distance between the statistics of the generated and the observed network, using the Earth mover's distance for all distributional statistics (path-length, component sizes, partner-count), and root-mean-square error for the remaining statistics (relatives count and percolation year).
For each statistic, we normalized the resulting distance measures to lie in the interval $[0, \, 1]$ and reported the mean, minimum, and maximum distances for each model and each statistic in Figure \ref{fig:homophily} panel E of the main text.

%%%%%%%%%%%%%%%%%%%%%%%%%%%%%%%%%%%%%%%%%%%%%%%%%%%%%%%%%%%%%%%%%%%%%%%%%%%%%%%%%%%%%%%%%%%%%%%%%%%%%
\section{Unnormalized distributional distances between generated and empirical network}\label{app:homophily_unnorm}
In Figure \ref{fig:homophily} panel E of the main text, we presented simulation evidence that homophily only plays a minor role in shaping the large-scale structure of family networks.
These results were normalized for ease of comparison across different summary statistics.
Here we present the unnormalized results.

\begin{figure}[h]
    \begin{center}
        \includegraphics[width=1\textwidth]{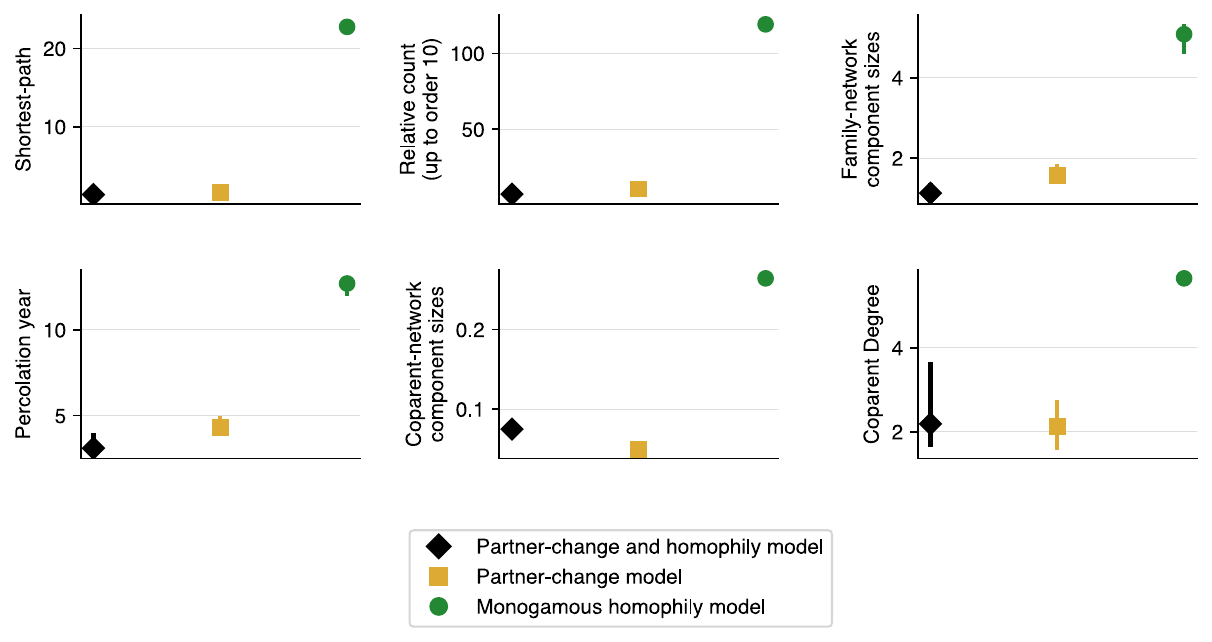}
    \end{center}
    \caption{
        Mean distance between structural summary-statistic distributions of generated and empirical network measured by Earth mover's distance~\cite{rubner2000earth} or root-mean-square error (see Appendix \ref{app:model_accuracy} for details)
        Vertical lines show the span between minimum and maximum distance observed across 10 independent simulations.
        Incorporation of both the observed partner-choice biases and partner-change behavior in our model results in surrogate networks that only resemble the Danish family network slightly more than we achieve when only accounting for partner-change behavior.
        Failing to account for partner-change behavior result in networks structurally very dissimilar from the Danish family network, even when accounting for partner-choice homophily.
    }\label{fig:homophily_unnorm}
\end{figure}

From Figure \ref{fig:homophily_unnorm} we see that our conclusion was not an artifact of the normalization: compared to the effect of partner-change behavior, partner-choice biases only play a minor role in shaping the emergent large-scale structure of family networks.

%%%%%%%%%%%%%%%%%%%%%%%%%%%%%%%%%%%%%%%%%%%%%%%%%%%%%%%%%%%%%%%%%%%%%%%%%%%%%%%%%%%%%%%%%%%%%%%%%%%%%
\section{Decision Margins and Reciprocity}
In our analysis of how individual behavior shapes large-scale family network structure, it is useful to distinguish between two different types of decisions, or \textit{decision margins}.
 
For an individual with a partner, the first decision margin concerns \textit{whether} an individual changes partners at all, which can be viewed as a binary decision.
An individual either remains with the current partner throughout their life, or at some point, they change partners and form a new partnership.
This aspect of behavior determines which individuals have the potential to create new links between otherwise unconnected family clusters.
 
The second margin addresses the question of \textit{whom} an individual partners with, given that a new partnership is formed.
Once an individual decides -- or has the opportunity -- to form a new partnership, there is a further decision regarding which partner to form a relationship with.
This choice is influenced by preferences or tendencies, such as similarities in age, education, or geographic origin.
 
This distinction between \emph{whether} and \emph{whom} is closely related to the concepts of the \textit{extensive} and \textit{intensive} margin, as commonly used in social science, particularly in economics and demography~\cite{heckman1993}.
The \emph{extensive margin} refers to the binary decision to change partners (i.e., participation or not), while the \emph{intensive margin} concerns the specific characteristics or qualities of the chosen partner, conditional on changing.
These margins have been used to clarify behavioral and policy effects in diverse contexts, including labor supply and fertility.
 
Yet, it is important to emphasize that the actual formation of a partnership is not determined by a single individual's choice alone.
Instead, partnership formation is inherently a mutual process: both individuals must agree to the relationship.
This mutual agreement is often referred to as a \textit{match}.
In other words, a new partnership can only be realized when both parties select each other, either actively or passively, from among the pool of possible partners.
Thus, when we refer to ``choice'' in the context of partnership, we are always describing this mutual process, not a unilateral selection.
 
This distinction is important for interpreting our analyses.
The structure of family networks is shaped both by the frequency with which individuals form new partnerships---which creates new potential connections in the network---and by the social or demographic characteristics that bring particular individuals together as matches.
 
In summary, our analyses separate the frequency of partnership changes (whether individuals change partners at all; the extensive margin) from the mutual selection process that determines who forms a new partnership with whom (the intensive margin).
This framework allows us to identify which aspects of partnership behavior most strongly influence the emergence of large-scale connectivity in family networks.

%%%%%%%%%%%%%%%%%%%%%%%%%%%%%%%%%%%%%%%%%%%%%%%%%%%%%%%%%%%%%%%%%%%%%%%%%%%%%%%%%%%%%%%%%%%%%%%%%%%%%%
%\section{Acknowledgments}
%The authors thank the anonymous reviewers for their valuable suggestions. 

%%%%%%%%%%%%%%%%%%%%%%%%%%%%%%%%%%%%%%%%%%%%%%%%%%%%%%%%%%%%%%%%%%%%%%%%%%%%%%%%%%%%%%%%%%%%%%%%%%%%%
%\section{Funding}

%%%%%%%%%%%%%%%%%%%%%%%%%%%%%%%%%%%%%%%%%%%%%%%%%%%%%%%%%%%%%%%%%%%%%%%%%%%%%%%%%%%%%%%%%%%%%%%%%%%%%
%\section{Author contributions statement}

%%%%%%%%%%%%%%%%%%%%%%%%%%%%%%%%%%%%%%%%%%%%%%%%%%%%%%%%%%%%%%%%%%%%%%%%%%%%%%%%%%%%%%%%%%%%%%%%%%%%%
%\section{Preprints}
%A preprint of this article is published at [DOI].

%%%%%%%%%%%%%%%%%%%%%%%%%%%%%%%%%%%%%%%%%%%%%%%%%%%%%%%%%%%%%%%%%%%%%%%%%%%%%%%%%%%%%%%%%%%%%%%%%%%%%
%\section{Data availability}

\end{document}